\documentclass[pra,twocolumn,showpacs,preprintnumbers,floatfix]{revtex4-1}
\usepackage{graphicx}
\usepackage{dcolumn}
\usepackage{bm}
\usepackage{latexsym,epsfig}
\usepackage{graphicx}
\usepackage{verbatim}
\usepackage{comment}
\usepackage{amsmath}
\usepackage{amssymb}
\usepackage{stmaryrd}
\usepackage{color}
\usepackage{grffile}
\usepackage[normalem]{ulem}
\DeclareGraphicsExtensions{.eps}

\newcommand{\beq}{\begin{equation}}
\newcommand{\eeq}{\end{equation}}
\newcommand{\bea}{\begin{eqnarray}}
\newcommand{\eea}{\end{eqnarray}}
\newcommand{\ben}{\begin{eqnarray*}}
\newcommand{\een}{\end{eqnarray*}}
\newcommand{\bfig}{\begin{figure}}
\newcommand{\efig}{\end{figure}}

\begin{document}
\title{Quantum phases of constrained dipolar bosons in coupled one-dimensional optical lattices}
\author{Manpreet Singh$^{1}$, Suman Mondal$^1$, B. K. Sahoo$^{2,3}$, Tapan Mishra$^1$ }
\affiliation{$^1$Department of Physics, Indian Institute of Technology, Guwahati-781039, India}
\affiliation{$^2$Atomic and Molecular Physics Division, Physical Research Laboratory, Navrangpura, Ahmedabad 380009, India}
\affiliation{$^3$State Key Laboratory of Magnetic Resonance and Atomic and Molecular Physics, Wuhan Institute of Physics and Mathematics, Chinese Academy of Sciences, Wuhan 430071, China}

\date{\today}

\begin{abstract}
We investigate a system of two- and three-body constrained dipolar bosons in a pair of one-dimensional 
optical lattices coupled to each other by the non-local dipole-dipole interactions. 
Assuming attractive dipole-dipole interactions, we obtain the ground state phase diagram of the system by 
employing the cluster mean-field theory. The competition between 
the repulsive on-site and attractive nearest-neighbor interactions between the chains yields three kinds 
of superfluids; namely the trimer superfluid, pair superfluid and the usual single particle superfluid along 
with the insulating Mott phase at the commensurate density. 
Besides, we also realize simultaneous existence of Mott insulator and superfluid phases for the two-
and three-body constrained bosons, respectively. We also analyze the stability of 
these quantum phases in the presence of a harmonic trap potential.
\end{abstract}


\maketitle

\section{Introduction}
Quantum phase transitions in strongly correlated quantum gases have been a topic of immense interest in the fields
of condensed matter and light-matter interactions, primarily due to the spectacular experimental progress in the field 
with the advent of 
modern technologies~\cite{blochrev,lewenstein_book}. Various interesting phases and phase transitions have been 
observed in the experiments using ultracold atoms 
in optical lattices and numerous theoretical predictions have been made in the last decade~\cite{lewenstein_book} 
following the prediction~\cite{fisher89,jaksch} and experimental observation of the superfluid(SF) to Mott insulator(MI) phase 
transition~\cite{bloch}. Inter-atomic interactions, which can be controlled by the laser 
intensity or by the technique of  
Feshbach resonances, play very important roles in such weakly interacting systems. 
In addition, lattice geometries contribute significantly 
for stabilizing novel quantum phases. Significant effects of multi-body interactions such as 
the three-body interaction have been realized in the experiment~\cite{swill}. 
Methods have been proposed to engineer the three-body interactions~\cite{daley,tiesinga,petrov}
which seem to have significant role on the ground state 
phase diagrams~\cite{daley_zoller,manpreet,manpreet_3bosl,greschner_3b}. 

On the other hand, long-range dipole-dipole interactions among atoms and molecules have also opened up  
new directions~\cite{pfaureview,baranovrev} in this field. The
non-local nature of these interactions have made it possible to realize quantum phases with long-range density wave order. The phase 
with density wave order when doped with extra particles or holes in the bosonic case stabilizes the exotic supersolid phase
~\cite{baranovrev,ferlaino,ketterless}. Moreover, long-range interaction may also lead to the Devil's stair case type structures
~\cite{baranovrev,sondhi,sansone}.

One of the remarkable properties of 
such off-site interactions is that these can be made repulsive as well as attractive by simply 
manipulating the directions of the dipoles by applying 
magnetic field. Due to the long-range nature of the interaction it is possible to 
 couple two non-local systems by suitably 
adjusting the associated dipole-dipole interactions between the constituents of both the systems. As a result 
 one can simulate interesting physics such as bosonic pair superfluid (PSF) - MI transition in bi-layer 
systems~\cite{lewenstein_bilayer}. In one dimension($1d$), analogous to this bi-layer geometry is 
 the system of two non-local chains coupled by the dipole-dipole 
interaction~\cite{santos1} which resembles a two-leg ladder.
Such low dimensional systems are very special to investigate the condensed matter physics 
due to active roles  played by quantum fluctuations~\cite{giamarchi_book}.
Quasi-$1d$ systems such as the two-leg ladder geometries can be engineered in the optical lattice experiment and manipulation of 
atomic species in such potentials gives rise to various interesting quantum 
phases~\cite{rigol_giamarchi_rmp,luthra,giamarchi1,danshita6to9,danshita1,manpreet_ladder1}. 

\begin{figure}[!b]
   \centering
\includegraphics[trim={0 8.8cm 9cm 4.9cm},clip,width=0.45\textwidth]{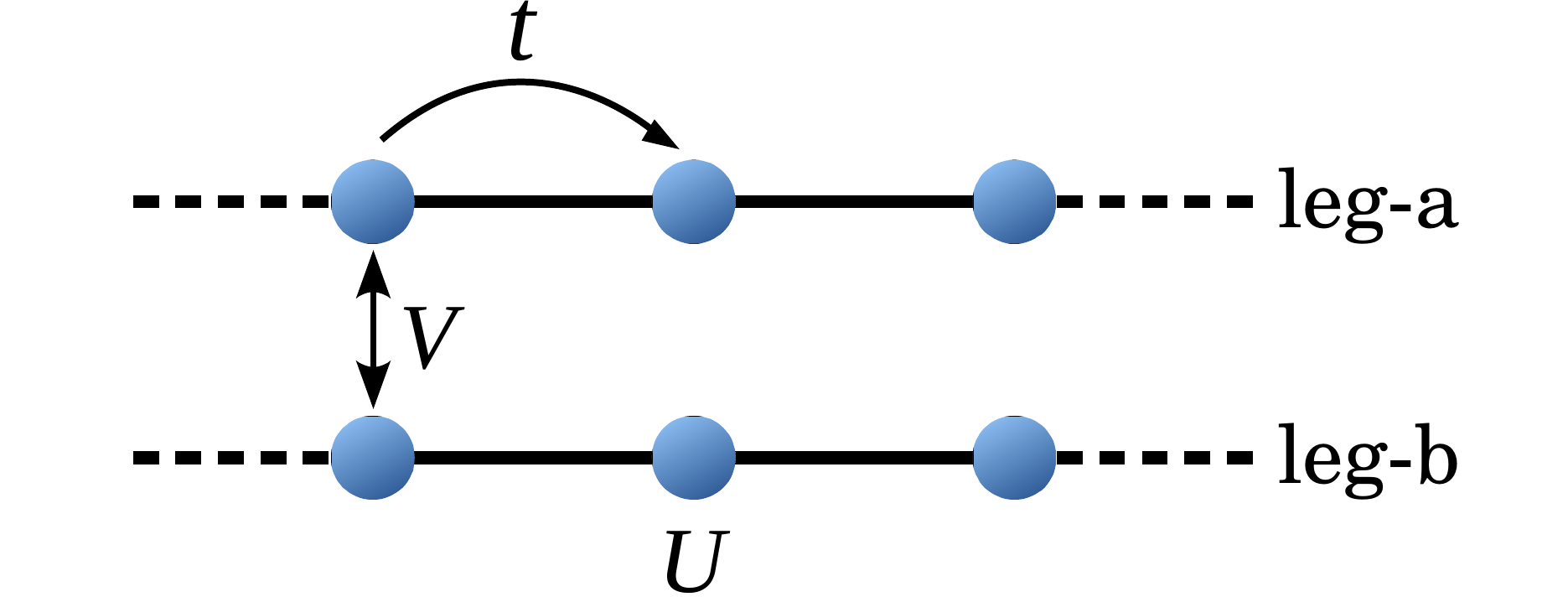}
\caption{(Color online) Two one dimensional optical lattices are coupled to each other by non-local interaction  
which is equivalent to a two leg ladder model. Leg-a and leg-b contain two- and three-body constrained bosons, respectively. 
The particles on 
both the chain interact among themselves with 
the non-local dipole-dipole interaction $V$ . The Hamiltonian for this system is given in Eq.\ref{eq:one}.}
\label{fig:fig1_ladder_fig}
\end{figure}

In this paper we consider a system of two spatially separated $1d$ optical lattices loaded with dipolar atoms as depicted in 
Fig.~\ref{fig:fig1_ladder_fig}. These kind of coupled systems resemble with 
the system of binary mixtures in optical lattices~\cite{santos1,miskin}. 
The system of multi-component atoms promises even richer platform to study novel quantum phase transitions. 
Bose-Bose and 
Bose-Fermi systems have been shown to exhibit various quantum phases due to the presence of the inter-species interactions 
along with the interactions 
within the individual species~\cite{altman,kuklov,santosbf,mishraps,mishrapscdw,mishra2ss,mathey,Titvinidze,Iskinmixture}. 
It has been shown that in Bose-Bose mixtures that atoms form superfluid of pairs of bosons called the 
pair superfluid (PSF) phase for attractive inter-species interactions~\cite{mathey}. 
On the other hand there occurs a spatial phase separation (PS) in the 
presence of a critical value of repulsive interaction~\cite{chen_wu,mishraps,ian}. 
Interesting quantum phases with composite fermions, where one fermion is associated with 
one or several bosons (bosonic holes) for attractive (repulsive) inter-species interaction~\cite{santosbf} have been proposed. 
Recent experimental observations of Bose-Bose and 
Bose-Fermi mixtures in optical lattices have paved the path to simulate such interesting physics 
in the laboratory~\cite{Esslinger,Ospelkaus,Best,Catani,Gadway,Taglieber}. 
Recently the dual MI phase in a system of Bose-Fermi mixture in Yb atoms has been observed~\cite{takahashidualmott}. 
We present a 
detailed discussion of different quantum phases in a wide parameter regime of the constrained dipolar bosons, which will also reflect the 
properties of systems of bosons and spin-polarized fermions in depth.

In this context we consider atoms in one chain that are hardcore in nature (i.e. the maximum occupation is one particle per site)
and in the other chain we impose three-body hardcore constraint
(i.e. maximum occupation is two particles per site). 
We also assume the atoms are arranged in such a way that they attract each other along the rung-direction of the ladder and there is 
no dipole-dipole interaction along the leg direction~\cite{santos1,manpreet_ladder1}. 
By using the self-consistent cluster mean-field 
theory, we analyze the ground state properties of this system and present various possible quantum phases that can 
arise due to the competition between the attractive dipole-dipole interactions and the on-site interactions. 

The remaining part of this paper is organized as follows. In Sec. \ref{sec:sec2}, we give details of the model and describe briefly
the method used in our calculations. We present and discuss our results in Sec. \ref{sec:sec3} before coming up with the concluding remarks in 
Sec. \ref{sec:sec4}.

\section{Model and method}
\label{sec:sec2}
The system considered here can be expressed in the framework of the extended Bose-Hubbard model (EBH) whose Hamiltonian is given by

\begin{eqnarray}
H=&-&t\sum_{i,\alpha=1,2}(a_{i,\alpha}^{\dagger}a_{i+1,\alpha}+H.c.)\nonumber\\
&+&{{U}\over{2}}\sum_{i,\alpha=1,2}[n_{i,\alpha}(n_{i,\alpha}-1)]\nonumber\\
&+&V \sum_{i}n_{i,1} n_{i,2}-\mu \sum_{i,\alpha=1,2}n_{i,\alpha}
\label{eq:one}
\end{eqnarray}
where ${a_{i}}^{\dagger} ({a_{i}})$ is the bosonic creation (annihilation)
operator at the site $i$ of leg-a or leg-b. Index $\alpha=1(2)$ represents leg-a (leg-b). The first term corresponds to 
hopping between the nearest neighbor sites within the same leg. Second and third terms describe the 
on-site and nearest-neighbor interactions between the legs of the ladder. Last term 
denotes the chemical potential. As mentioned before we consider that the leg-a of the 
ladder consists of only hard-core bosons (HCB)($(a^\dagger)^2=0$) and leg-b 
contains three-body constrained bosons (TCB)($(a^\dagger)^3=0$).
On top of that the bosons in leg-b possess repulsive on-site interaction $U$, while the dipole-dipole interaction $V$ is attractive.

We perform our studies using the self consistent cluster mean-field theory (CMFT) approximation which has been used extensively in the recent 
years. This method takes care of non-local 
correlations and hence, it is more powerful as compared to the single site mean-field theory. The accuracy depends on the 
size of the cluster~\cite{daniel,danshitacmft,hassan,manpreet_ladder1,luhman}. Since we consider  
attractive non-local interaction in the present study, 
single site mean-field theory cannot encapsulate the underlying physics anticipated in this system. 
In principle, it would be appropriate to consider a more sophisticated method like the density matrix renormalization 
group (DMRG) method \cite{white} or the Matrix Product States (MPS) approach \cite{scholwoek}, 
which are extremely apt to obtain the ground state properties of 
low dimensional systems. However, it can be understood from the following analysis that the advantage of 
considering the CMFT method is that it captures the 
essential physics to probe the intended signatures adequately with very less computing power.

\section{Results and Discussion}
\label{sec:sec3}
\begin{figure}[b]
   \centering
\includegraphics[width=0.45\textwidth]{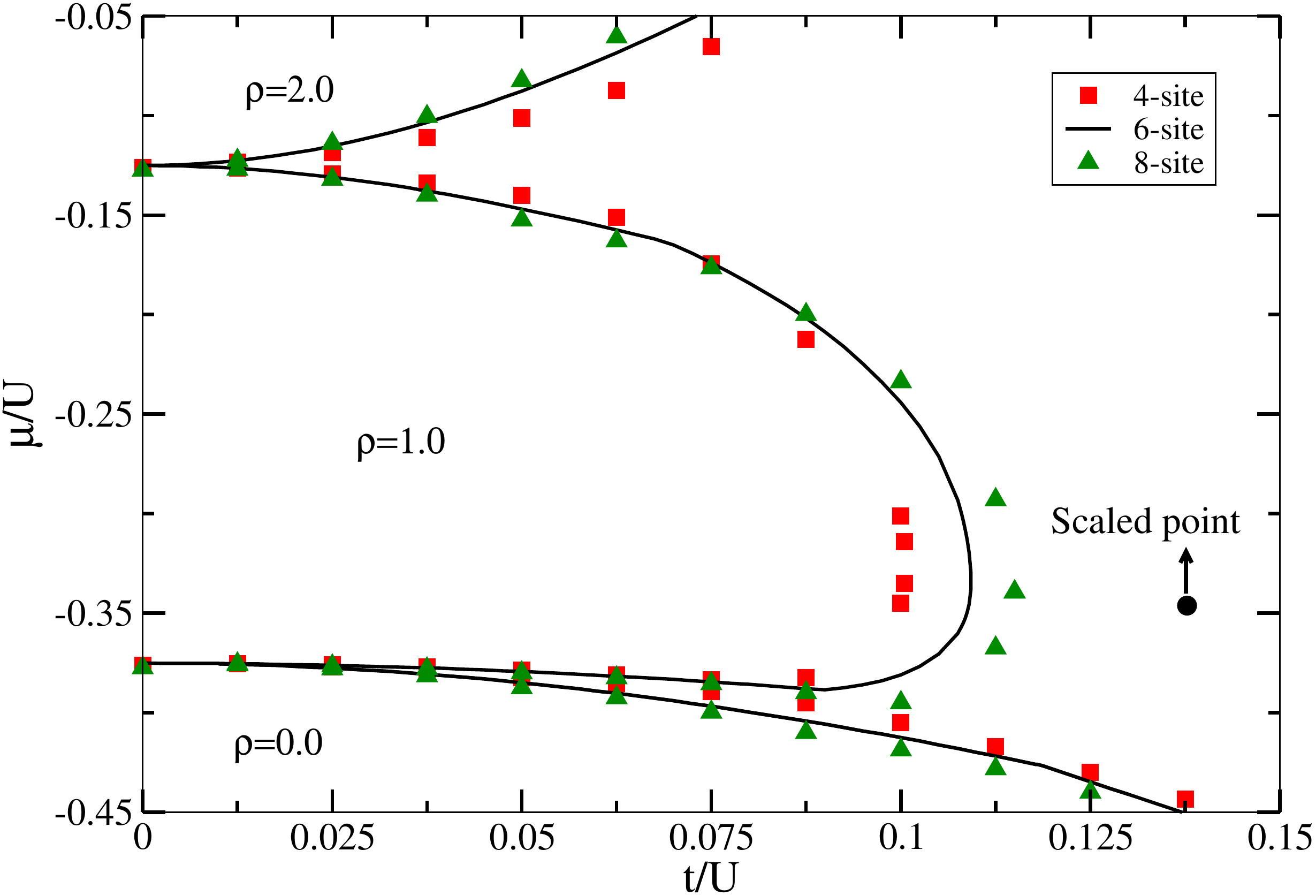}
\caption{(Color online)Phase diagram for TCBs in both the chains for $|V|/U=−0.75$ in the $t/U$ -$\mu/U$ 
plane as considered in Ref.~\cite{santos1}. Red squares, green up-triangles and the black solid curve represent 
the phase boundary for 4-, 6- and 8-sites cluster respectively. 
Scaled critical 
point for MI-SF transition $(t/U)_c(=0.1377)$ is shown by a solid black circle.}
\label{fig:fig2_tvar_U40_V-30}
\end{figure}

Before proceeding further, we first validate our CMFT method by carrying out calculations of the already existing results for a limiting case. 
The system of two decoupled chains, which are connected only through the non-local dipole-dipole interaction $V$, has been 
studied earlier using strong coupling expansion and the MPS method~\cite{santos1}.
It was found that the ground state phase diagram exhibits a direct MI-PSF transition as a function of 
$t/U$ for large values of $|V|$. 
Due to the attractive interaction $V$, a pairing takes place between the bosons on the two different legs of the ladder
at incommensurate densities while at commensurate density, the MI phase appears in both the chains. 
The lower boundary of the MI lobe gets distorted 
with an increase in the value of $|V|$. As the strength of $|V|$ increases further, the lower Mott boundary first gets flattened out
and then starts bending near the tip of the lobe. 
The Mott lobe in this case is distorted as a function of hopping and a re-entrant type behavior appears in the phase diagram.
This behavior is 
also clearly visible from another calculation performed using the DMRG method~\cite{danshita2}. 

\begin{figure}[!t]
   \centering
\includegraphics[width=0.45\textwidth]{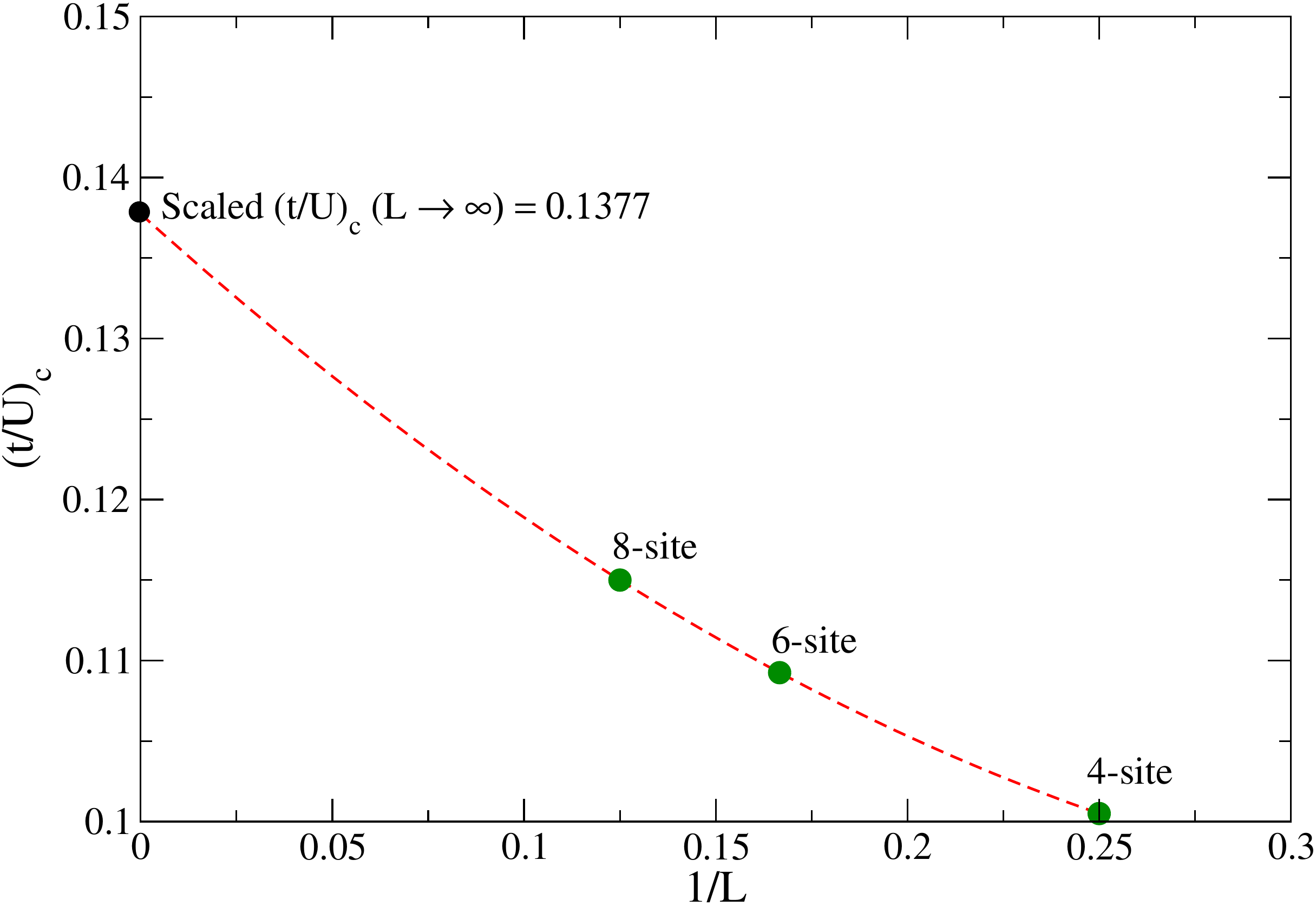}
\caption{(Color online) Scaling of the MI-SF transition critical point $(t/U)_c$ with respect to the different cluster sizes
at $|V|/U=−0.75$. Green solid circles represent the $(t/U)_c$ values for respective cluster sizes, red dashed lines represents
the scaling and the black solid circle on Y-axis denotes the value of $(t/U)_c$ scaled to thermodynamic limit $(=0.1377)$.}
\label{fig:fig3_scaling}
\end{figure}

To reproduce the above findings, we consider exactly the same parameters as considered in Ref.~\cite{santos1} 
and employ the CMFT approach to obtain the MI-PSF phase boundary as shown in Fig.~\ref{fig:fig2_tvar_U40_V-30}. 
The black solid curve shows the MI-PSF
phase boundary for a six-site cluster and $|V|/U=0.75$. This already shows the bending of curvature of the 
lower boundary of the Mott lobe as predicted 
in Ref.~\cite{santos1}. In order to affirm these result more distinctly we perform calculations with 8-site cluster which 
are shown by the green triangles.   We find that the tip of the Mott lobe approaches towards the value obtained using the MPS method.  
Further, we perform a cluster size extrapolation to estimate the critical value $(t/U)_c$ at the tip of the Mott lobe as shown in 
Fig.~\ref{fig:fig3_scaling}. The extrapolated point is shown as the black filled circle in Fig.~\ref{fig:fig2_tvar_U40_V-30}.
The finite size extrapolation leads to $(t/U)_c\approx0.14$ against the value $(t/U)_c\approx0.18$, which 
was shown in Ref. ~\cite{santos1}. The region depicted by $\rho=0$ and $\rho=2$ 
correspond to the empty and full states respectively. It is obvious from the above analysis 
that our CMFT method is able to predict the ground state of the system of TCBs 
reasonably well and can be used to 
perform a ground state analysis of the aforementioned model described by  Eq.~(\ref{eq:one}).

We now proceed on to discuss the main results obtained for the case, when one of the 
linear chains is occupied by the HCBs and the other chain 
by the TCBs. 
This situation can also mimic a two leg ladder where one leg of the ladder contains only HCBs (say, leg-a) 
and the other leg contains only TCBs (say, leg-b). 
Like previous case these bosons are also considered to be dipolar in nature and 
the dipole orientation is such that both the legs are coupled via attractive dipole-dipole  
interaction $V$ and there is no dipole-dipole interaction along the legs. 
In addition to this, the TCBs interact through a finite on-site interaction term $U$. Henceforth, $U$ will 
be used to denote the on-site two-body interaction strength for TCBs only (as for HCB, $U\rightarrow\infty$). 
We consider two scenarios in the following subsections. In the first case $U = 0$ and in the second case we set $U \neq 0$.

\begin{figure}[!b]
   \centering
\includegraphics[width=0.45\textwidth]{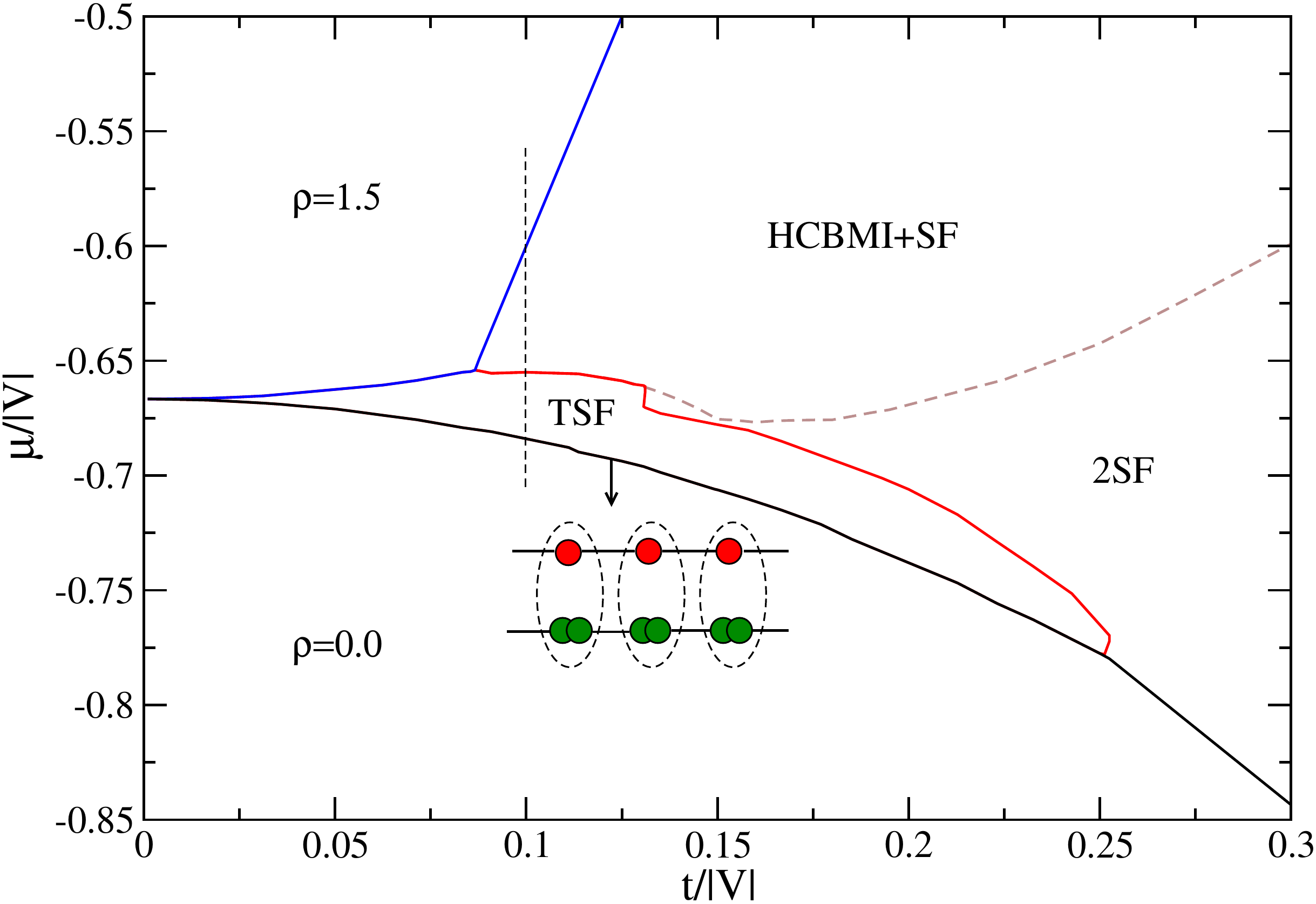}
\caption{(Color online) Phase diagram for model given in Eq.\ref{eq:one} in $\mu/|V|$ vs. $t/|V|$ plane, 
with leg-a (leg-b) containing HCB (TCB with $U=0$) with varying $t/|V|$.}
\label{fig:fig4_HC+3b_t1_U0_Vvar_phase-diagram}
\end{figure}

\subsubsection{\bf \large TCBs with U=0}

First we investigate the effect of an attractive $V$ on the system in the absence of two-body interaction between the TCBs, i.e. $U=0$.
The detailed phase diagram for this case is shown in Fig.~\ref{fig:fig4_HC+3b_t1_U0_Vvar_phase-diagram}.
As the on-site repulsion is absent, the bosons on different legs form a bound state of one HCB and two TCBs (HCB+2TCBs) when 
$|V|$ is sufficiently large compared to $t$. 
This happens due to the obvious reason as the two- and three-body constraints prevent
more than one and two atoms on a single site of leg-a and leg-b, respectively. When the density is small, these bound state can move 
freely in the chains giving rise to a SF phase, which we call a trimer superfluid (TSF) phase. The TSF 
phase is depicted in the cartoon as the bound state of one HCB (green circle) and two TCBs (red circles). 
For small to moderate values of $t/|V|$, 
the TSF phase is always present but when $t/|V| \gtrsim 0.25$, 
there is a direct transition from vacuum to 2SF phase.  For intermediate values of 
$t/|V|$, when the density of bosons and hence the value of $\mu/|V|$  increases, at 
some point leg-a becomes a Mott insulator with one atom in every site due to the hardcore constraint. 
In this limit we call this as the hardcore boson MI (HCBMI) phase. At the same time the leg-b shows a SF signature, as the trimer which was formed 
before can not move in the limit when the leg-a is full. We call this region of the phase diagram as the HCBMI+SF phase. 
Further increase in density leads to the saturation at $\rho=1.5$, i.e. $\rho=1$ and $2$ for HCBs and TCBs respectively. 
However, for small values of $|V|$ the system first goes to a 2SF phase from the TSF phase and then to the HCBMI+SF phase before saturating. 
For very large values of $|V|$ the system goes directly into the saturation from the TSF phase with increase in density of particles.  
In Fig.~\ref{fig:fig4_HC+3b_t1_U0_Vvar_phase-diagram} , the  HCBMI+SF phase is separated from the 2SF phase by the dashed line.

\begin{figure}[!t]
   \centering
\includegraphics[width=0.45\textwidth]{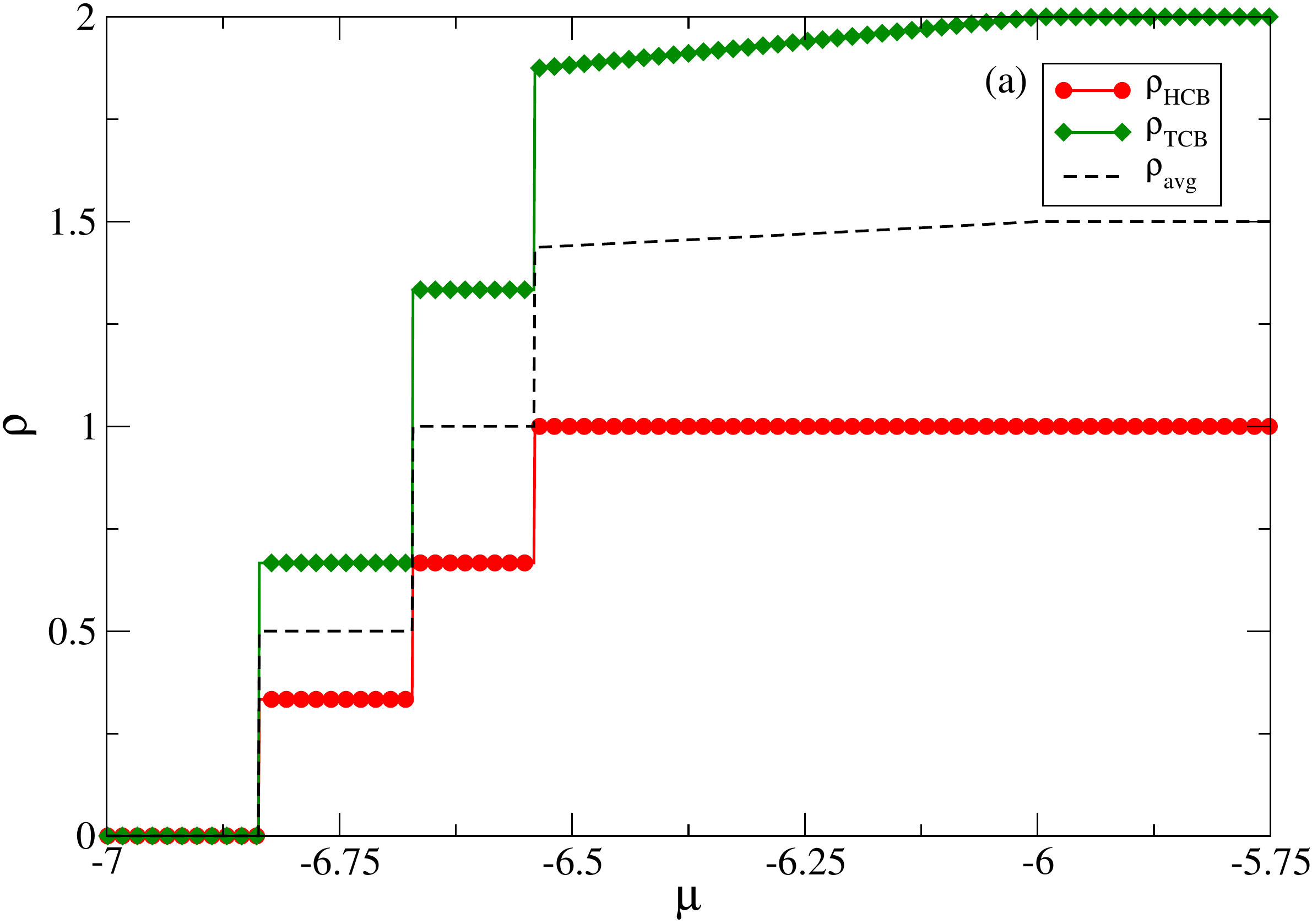}
\includegraphics[width=0.45\textwidth]{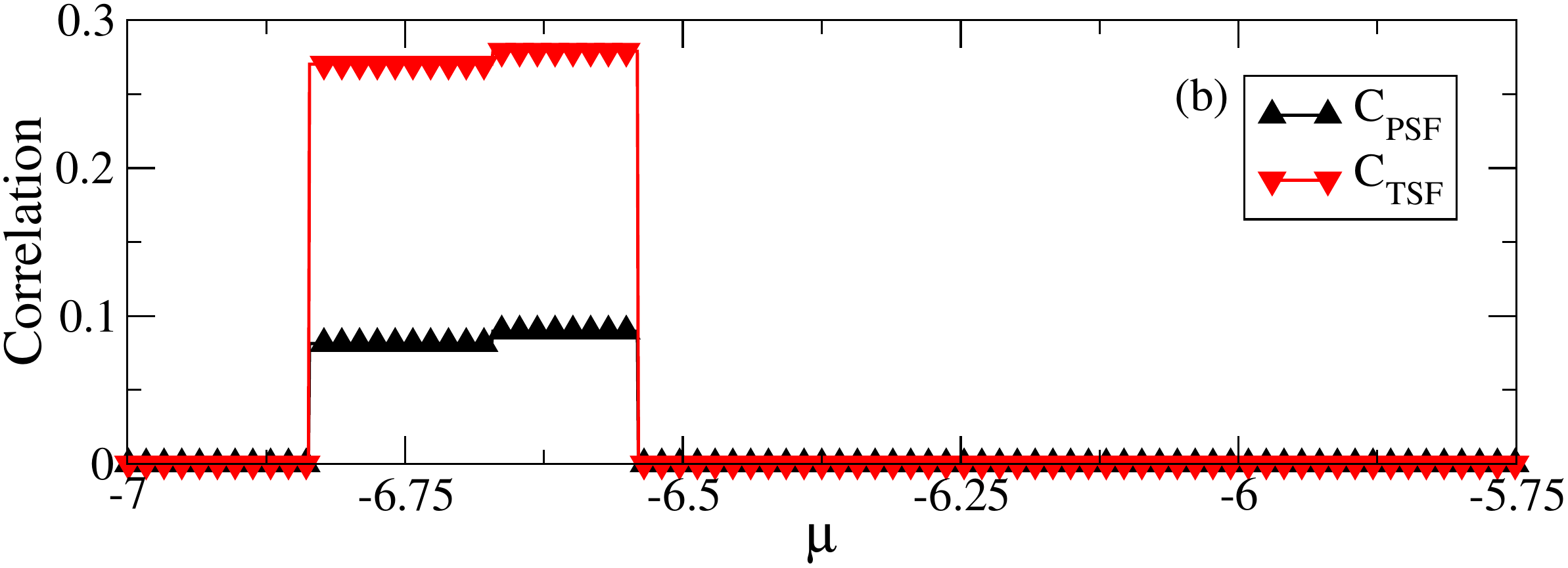}
\includegraphics[width=0.45\textwidth]{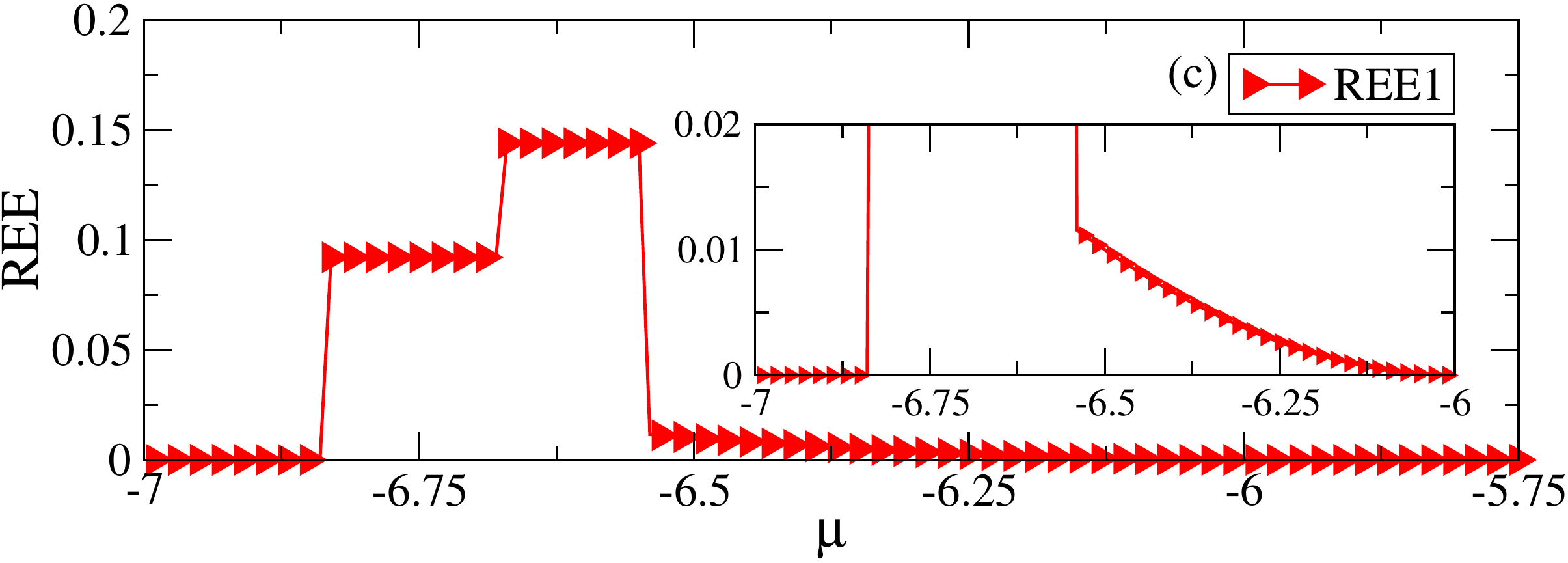}
\caption{(Color online)(a)$\rho$ vs. $\mu$
(b)Dimer and trimer correlations and
(c)R\'{e}nyi entropy 
corresponding to a vertical cut along $U/|V|=0.0$ in Fig.\ref{fig:fig8_HC+3b_Uvar_V-10_phase-dig}.
Inset shows the zoomed in region in which REE is finite at first (HCBMI+ASF) and then gradually drops to 0 as the
system approaches saturation.}
\label{fig:fig10}
\end{figure}

To obtain the phase diagram for this case we analyze the densities and the superfluid order parameter $\phi$ of 
the two legs individually as well as of the whole system for different values of $t/|V|$ as a function of the 
chemical potential $\mu$ in a 6-site cluster (3 sites each in leg-a and leg-b).
The $\rho$ vs. $\mu$ plot across a cut in the phase diagram (dashed vertical line in 
Fig.~\ref{fig:fig4_HC+3b_t1_U0_Vvar_phase-diagram}) for $t/|V|=0.1$ is shown in Fig.~\ref{fig:fig10}(a).
In Fig.~\ref{fig:fig10}(a) the HCB, TCB and average density of the system are shown by red circles, green diamonds
and black-dashed lines, respectively.
It can be seen from this figure that there appears several discontinuous jumps in the densities as a function of $\mu$. 
These step-wise 
jumps correspond to the formation of bound states. It is to be noted that in the CMFT approach, the 
formation of bound states can be inferred from these discrete jumps in density for creation of every bound state. This has been 
confirmed in our earlier work~\cite{manpreet_ladder1}.
As $\mu/|V|$ increases both the legs, hence the system starts filling up. In the case of trimer formation, 
the filling pattern is such that 
for every single particle in leg-a, there are two particles in leg-b. 
As we go on increasing $\mu$ more particles are introduced and more such bound states are formed. 
This process continues until 
both the legs are fully occupied and the system attains its maximum possible density ($\rho=1.5$).
The stepwise jump in the case of HCBs at 
$\rho=0.33, 0.66$ and $1.0$ corresponds to 1-, 2- and 3-particle states, respectively, of the leg-a.
Similarly, the stepwise jump for TCBs at $\rho=0.66, 1.33$ and $2.0$ indicate 2-, 4- and 6-particle states, respectively, of leg-b. 
The average density of the system also behaves like wise.

\begin{figure}[!b]
   \centering
\includegraphics[width=0.45\textwidth]{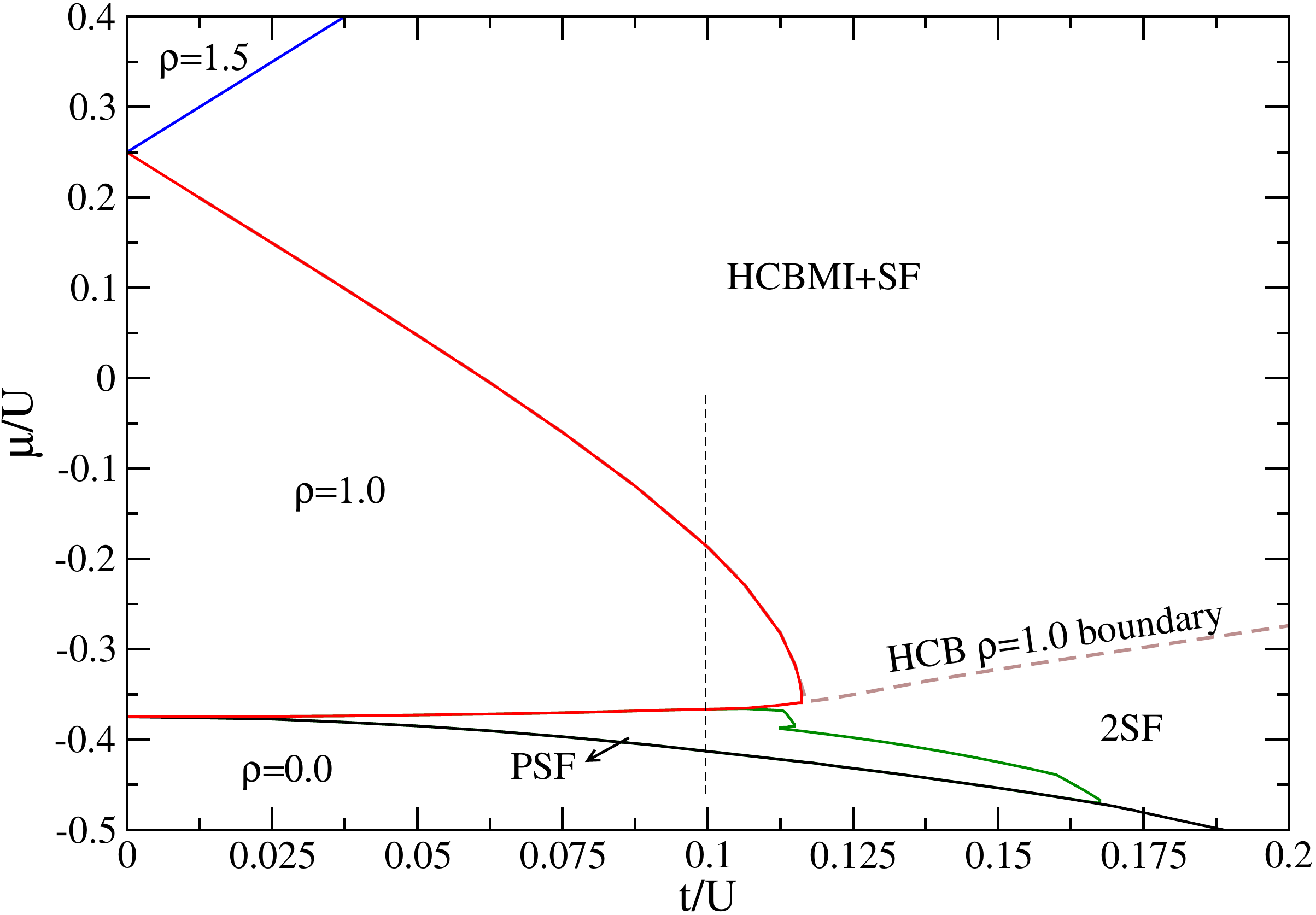}
\caption{(Color online) Phase diagram for model given in Eq.\ref{eq:one} in $\mu/U$ vs. $t/U$ plane, 
with leg-a (leg-b) containing HCB (TCB with finite two-body interaction).}
\label{fig:fig6_HC+3b_U40_V-30_tvar_PD}
\end{figure}

\begin{figure}[!t]
   \centering
\includegraphics[width=0.45\textwidth]{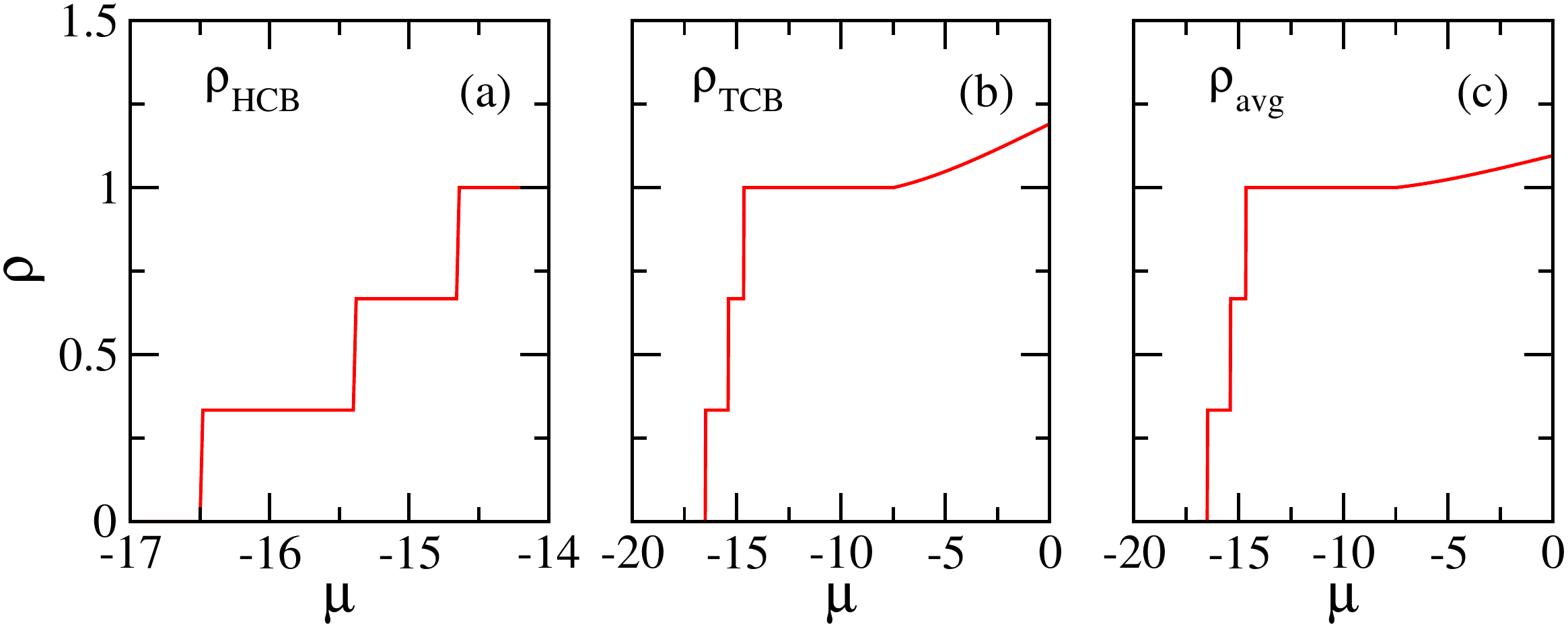}
\caption{(Color online) $\rho$ vs. $\mu$ plot corresponding to the cut shown by dashed line across the phase diagram shown 
in Fig.\ref{fig:fig6_HC+3b_U40_V-30_tvar_PD}.}
\label{fig:fig7_HC+3b_U40_V-30_rho-mu}
\end{figure}
\subsubsection{\bf \large TCBs with  U$\neq$ 0}
We now discuss the situation for finite on-site two-body interaction $U$ between the TCBs. 
In this case the presence of on-site repulsion will play an important role as it will break the trimer phase and stabilize the 
MI and PSF phases. As mentioned before, in the case of soft-core bosons in both the legs 
it has been shown that there exists a MI-PSF phase transition as a function of $(t/U)$ for finite values of $|V|$ keeping the 
ratio $U/|V|=1.33$ (equitably $|V|/U=0.75$ as considered in Ref.~\cite{santos1}. 
The PSF phase is the bound state of two bosons each from different legs. 
In the case of constrained bosons studied in this case, we show that the 
MI-PSF transition also occurs for large values of $|V|/U$. However, the phase diagram in this case shows interesting features due to 
the two- and three-body constraints. We vary the ratio $(t/U)$, while keeping $V$ fixed to find out the 
existence of various phases. The phase diagram obtained for this case is shown in Fig.~\ref{fig:fig6_HC+3b_U40_V-30_tvar_PD}. 
Compared to the previous case of TCBs with no local two-body interaction, here we do get an MI phase when the densities of both
the legs becomes unity. However, before entering the MI phase the system undergoes a transition from vacuum to a
PSF phase. The existence of this PSF phase can be understood with the help of $\rho$ vs. $\mu$ plot given in 
Fig.~\ref{fig:fig7_HC+3b_U40_V-30_rho-mu} along the dashed vertical line for $t/U=0.1$. 
For an attractive $V$, the system favors a bound state between the bosons of the two legs.
If both the legs contain one-particle each a pair is formed. With increase in the chemical potential the number of particle in 
the system increases. As the system favors the formation of bound pairs, the number of particles in the system increase in steps of 
two particles at a time. 
This results in the step wise jump in the $\rho$ vs. $\mu$ plot as discussed before. The densities of 
HCBs, TCBs as well as the average density of the system is plotted in Fig.~\ref{fig:fig7_HC+3b_U40_V-30_rho-mu}(a), (b) and (c) 
respectively for a 6-site cluster. The jumps in all the three plots signify the existence of the PSF phase. 
The plateau at $\rho=1$ 
corresponds to the MI phase and the shoulder above the plateau is the gapless SF phase. 
When the on-site interaction strength $U$ is small then the system enters into a 2SF phase with the chemical potential.
However, for large values of $U$ the system favors an MI phase at commensurate density of both the 
legs where the density of both the legs are equal to unity. 
Further increasing the chemical potential leads to the HCBMI+SF phase for all values of $t/U$ 
as discussed in the previous section. The system saturates at $\rho_{avg}=1.5$. 
It is to be noted that the PSF phase, which appears along the top boundary of the MI lobe, 
disappears due to the hardcore constraint in one leg. This prevents the motion of the pairs once leg-a is full. As a result the 
bosons in leg-b can move freely in this region and hence, it is responsible for breaking the pair formation. 
On the other hand the re-entrant type behavior of 
the lower MI lobe disappears in this case owing to the fact that the hopping process for the HCBs is restricted in the MI phase and 
the Mott boundaries are not quadratic anymore~\cite{santos1}. 

\begin{figure}[!b]
   \centering
\includegraphics[width=0.45\textwidth]{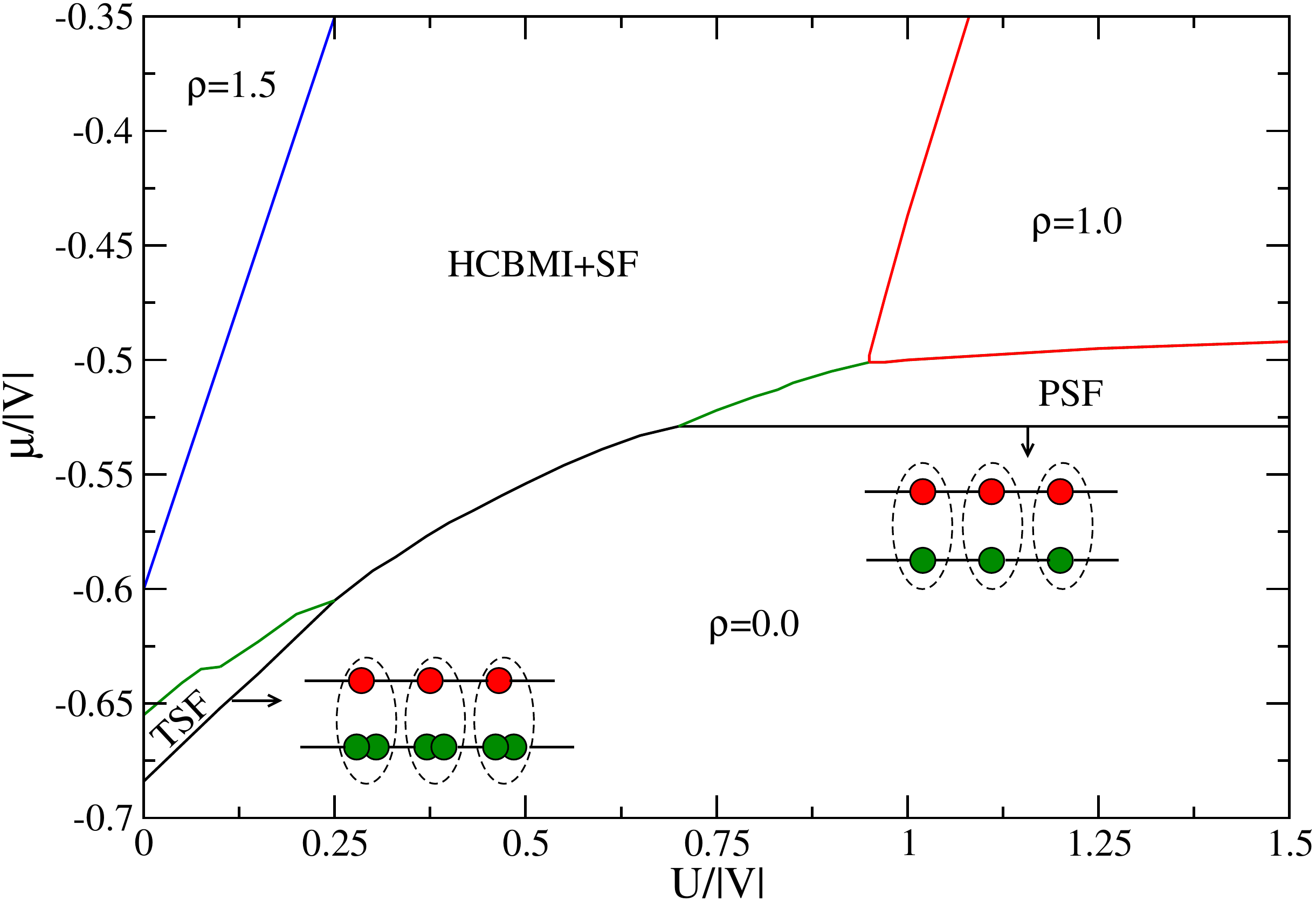}
\caption{(Color online) Phase diagram for model given in Eq.\ref{eq:one} in $\mu/U$ vs. $t/U$ plane, 
with leg-a (leg-b) containing HCB (TCB with finite two-body interaction) at $V=-10.0$.}
\label{fig:fig8_HC+3b_Uvar_V-10_phase-dig}
\end{figure}

\begin{figure}[!b]
   \centering
\includegraphics[width=0.45\textwidth]{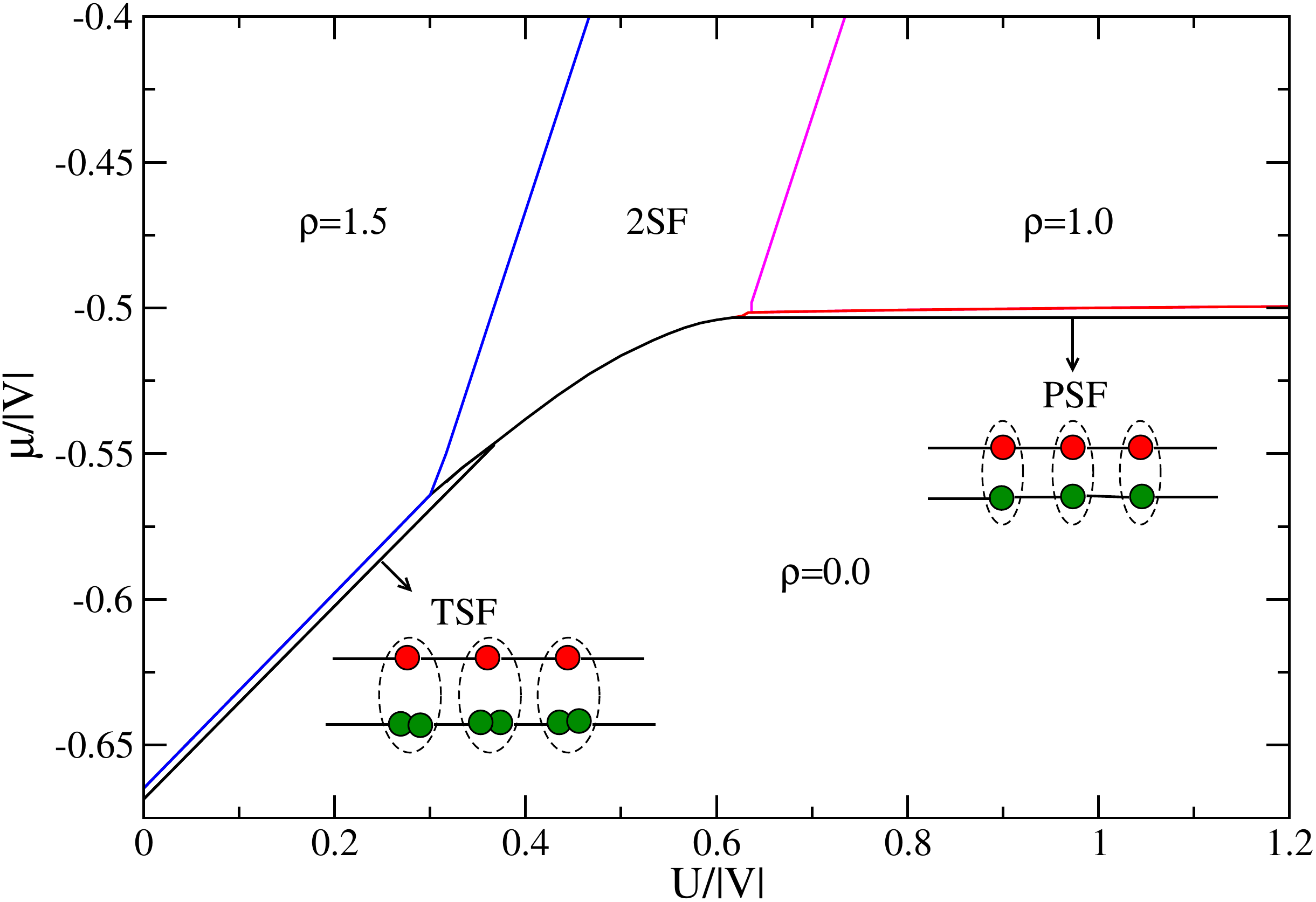}
\caption{(Color online) Phase diagram for model given in Eq.\ref{eq:one} in $\mu/|V|$ vs. $U/|V|$ plane, 
with leg-a and leg-b containing HC and three-body constrained bosons for $t/|V|=0.03$.}
\label{fig:fig9_HC+3b_full-phase-dig}
\end{figure}

\begin{figure}[!b]
   \centering
\includegraphics[width=0.45\textwidth]{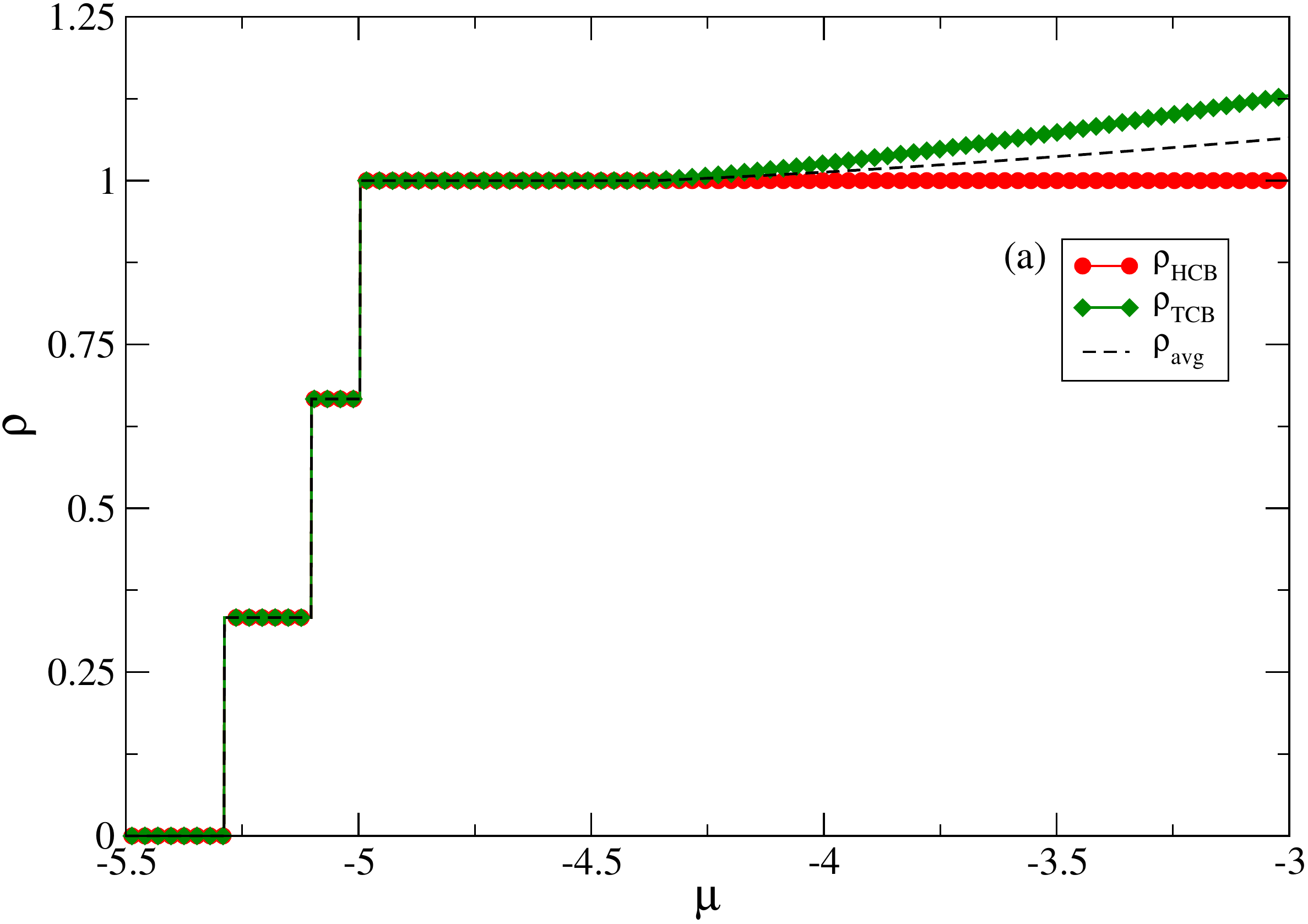}
\includegraphics[width=0.45\textwidth]{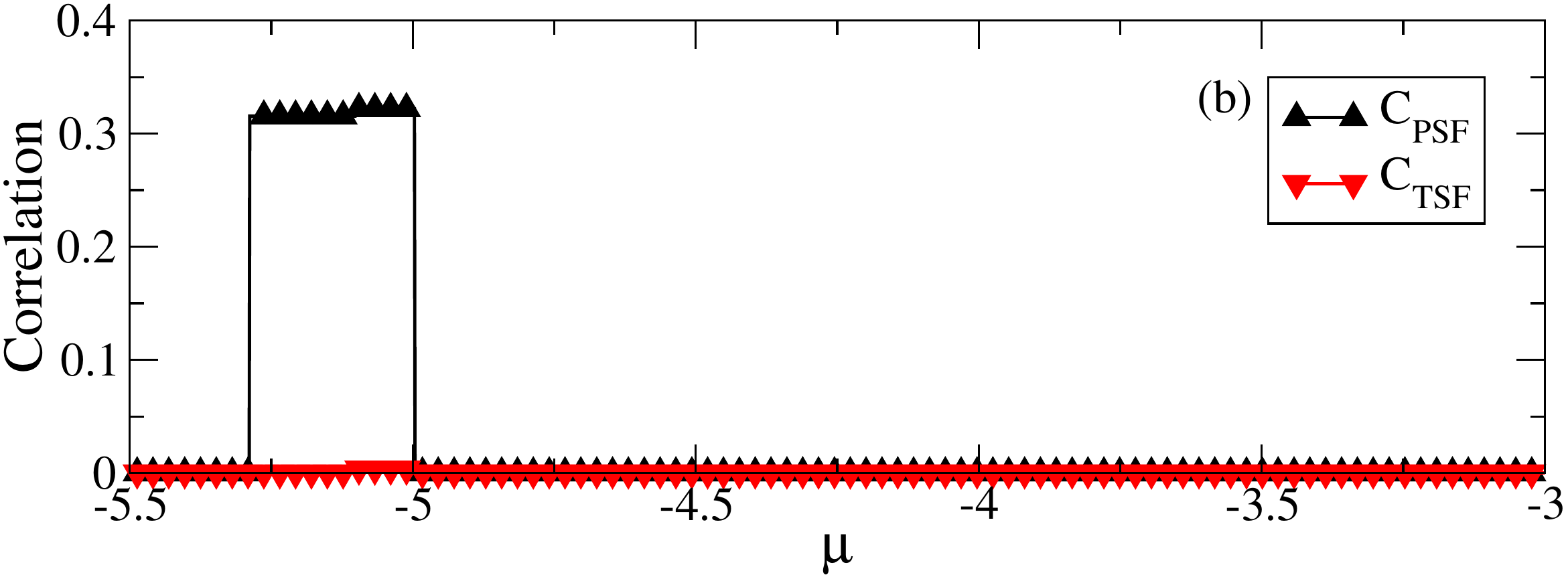}
\includegraphics[width=0.45\textwidth]{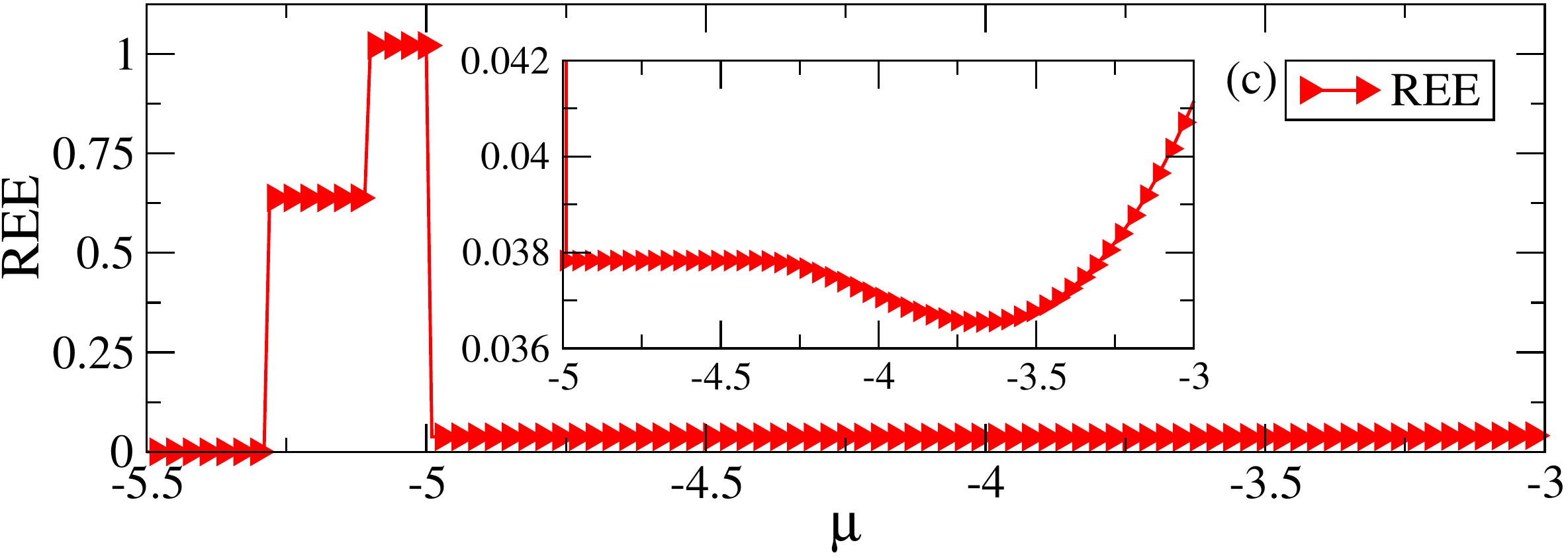}
\caption{(Color online) (a)$\rho$ vs. $\mu$
(b)Dimer and trimer correlations and
(c)R\'{e}nyi entropy 
corresponding to a vertical cut along $U/|V|=1.0$ in Fig.\ref{fig:fig8_HC+3b_Uvar_V-10_phase-dig}.
Inset shows the zoomed in region where there is a phase transition from MI to HCBMI+SF phase
and a change in slope of REE is visible.}

\label{fig:fig11}
\end{figure}

After discussing two cases above it is informative to perform a careful analysis of the 
competition between repulsive $U$ and attractive $V$, which may reveal a more detailed phase diagram. In this regard, we 
explore the system by varying the $U/|V|$ ratio for a large range of values. We 
investigate how the phase diagram changes when the on-site interaction $U$ of 
TCBs in leg-b is changed for a finite value of inter-leg interaction $V$. 
First, we discuss the case for small $|V|$. We fix $t/|V|=0.1$ and 
vary $U/|V|$ to obtain the phase diagram, which is shown in Fig.~\ref{fig:fig8_HC+3b_Uvar_V-10_phase-dig}. 
When $U/|V|=0$, this corresponds to the 
case when the system exhibits the trimer phase as discussed in the previous section. 
Further increase in the value of $U/|V|$ does not allow for the formation of the trimer phase 
as it prevents the two TCBs to occupy the same site 
in leg-b. Therefore, the TSF phase survives for a range of values of $U/|V|$ between $0.0$ and $0.25$.
Following which we only see HCBMI+SF phase without any signature of a paired phase or the MI phase. 
For a relatively larger value of $U/|V|$ (beyond $U/|V|=0.75$),
the density in both the legs are same. As $|V|$ is still finite, a dimer formation occurs between the 
legs at densities $0.33$ and $0.66$. This is indicated in the $\rho$ vs. $\mu$ plot for $U/|V|=1.0$ in Fig.~\ref{fig:fig11}(a). 
The particle number in both the legs always jumps in steps of one until the system goes to the HCBMI+ASF phase 
(for $0.75\lesssim U/|V|< 1.0$) or the MI phase (for $U/|V|\gtrsim 1.0$).

Now we discuss the case for large $|V|$. As before, we obtain the phase diagram by varying $U/|V|$ while keeping $t/|V|=0.03$,
which is shown in Fig.~\ref{fig:fig9_HC+3b_full-phase-dig}. The overall phase diagram 
and phases are similar as found in the previous case.
A major difference is seen in the width and phase boundaries of the TSF and the PSF regions. 
It is seen that the TSF and PSF phase shrink
and get reduced to a very small region.
For intermediate values of $U/|V|(\approx0.4-0.6)$, the repulsion between TCBs is strong enough to break the pairing in the TSF
phase. In this range of $U/|V|$ there is a direct transition from vacuum to the SF phase and then to a 
fully occupied state. As $U/|V|$ is increased further, the system exhibits a PSF phase for incommensurate densities as 
$V$ is still finite and attractive in nature. At integer densities there is a transition to the MI phase in which
both the HCBs and the TCBs have average densities equal to unity. 
When in the MI phase, as $\mu$ is increased, the TCBs drive the system
first into a SF phase and eventually saturates at $\rho=1.5$.

To confirm the existence of various phase transitions discussed above we calculate correlation functions as well as 
the R\'{e}nyi entanglement entropy (REE) of the system. 
We calculate the pair correlation function as 
\begin{equation}
 C_{PSF}=\langle{a_i}^\dagger{b_i}^\dagger{a_j}{b_j}\rangle,
 \end{equation}
 and the trimer correlation function as 
\begin{equation}
 C_{TSF}=\langle{a_i^ \dagger}(b_i^\dagger)^2 a_j  (b_j)^2\rangle .
\end{equation}
to infer the signatures of the PSF and the TSF phases, respectively.
These quantities are plotted with respect to the chemical potential $\mu$ in Fig.~\ref{fig:fig10}(b) and Fig.~\ref{fig:fig11}(b) 
corresponding to the cuts shown in 
Fig.~\ref{fig:fig4_HC+3b_t1_U0_Vvar_phase-diagram} and Fig.~\ref{fig:fig6_HC+3b_U40_V-30_tvar_PD}, respectively. 
It can be seen from Fig.~\ref{fig:fig10}(b) that $C_{TSF}$(red triangles) clearly dominates $C_{PSF}$
in the region where the system exhibits a trimer phase as shown in Fig.~\ref{fig:fig10}(a). Both the correlation functions are zero 
in the HCBMI+SF phase. On the other hand, $C_{PSF}$ is larger than $C_{TSF}$ in the dimer phase as shown in Fig.~\ref{fig:fig11}(b). 

We further calculate the REE to complement our findings in this work. This is an important quantity
to probe quantum phase transitions in optical lattices which has been measured recently in an experiment 
on a Bose-Hubbard system \cite{greiner_2015}. The REE of 
$n^{th}$ order is defined as:
\begin{equation}
 S_n(A)={{1}\over{1-n}}~ log~ Tr({ \hat{\rho} _{A}}^n)
\end{equation}
where
${\hat{\rho}_{A}}$ is the reduced-density matrix of a subsystem A entangled with it's complement B.
For our calculations we focus only on the $2^{nd}$ order REE which can then be written as 
$S_2(A)=log~ Tr({ \hat{\rho} _{A}}^2)$.
The system considered can be divide into two subsystems as follows: 
Four left most sites can form one subsystem and the remaining (right most two sites) form another subsystem.
We calculate the REE in this configuration for the cuts along $U/|V|=0.0$ and $1.0$ in the phase diagram shown in 
Fig.~\ref{fig:fig8_HC+3b_Uvar_V-10_phase-dig} and plot it as a function of $\mu$ in Fig.~\ref{fig:fig10}(c) and 
\ref{fig:fig11}(c), respectively. 
As expected we observe finite REE in the SF, PSF and TSF phases. 
However in the gapped MI phase and in the saturated region, the REE reduces considerably. 
In Fig.~\ref{fig:fig10}(c), it can be seen that the REE is finite in the TSF region and as the system 
moves into the HCBMI+SF phase there is a significant change in REE which indicates a phase transition. As there
is a contribution from SF phase REE is still finite, which can be seen in the inset of Fig.~\ref{fig:fig10}(c).
As the system approaches saturation, REE also reduces to 0.
Similar features are seen for the transition from the 
PSF to MI then to HCBMI+ASF phases as shown in Fig.~\ref{fig:fig11}(c). 
Our findings from the REE calculations
are therefore consistent with the phase transitions indicated by the corresponding $\rho$ vs. $\mu$ and correlation function plots.

\begin{figure}[!b]
   \centering
\includegraphics[width=0.45\textwidth]{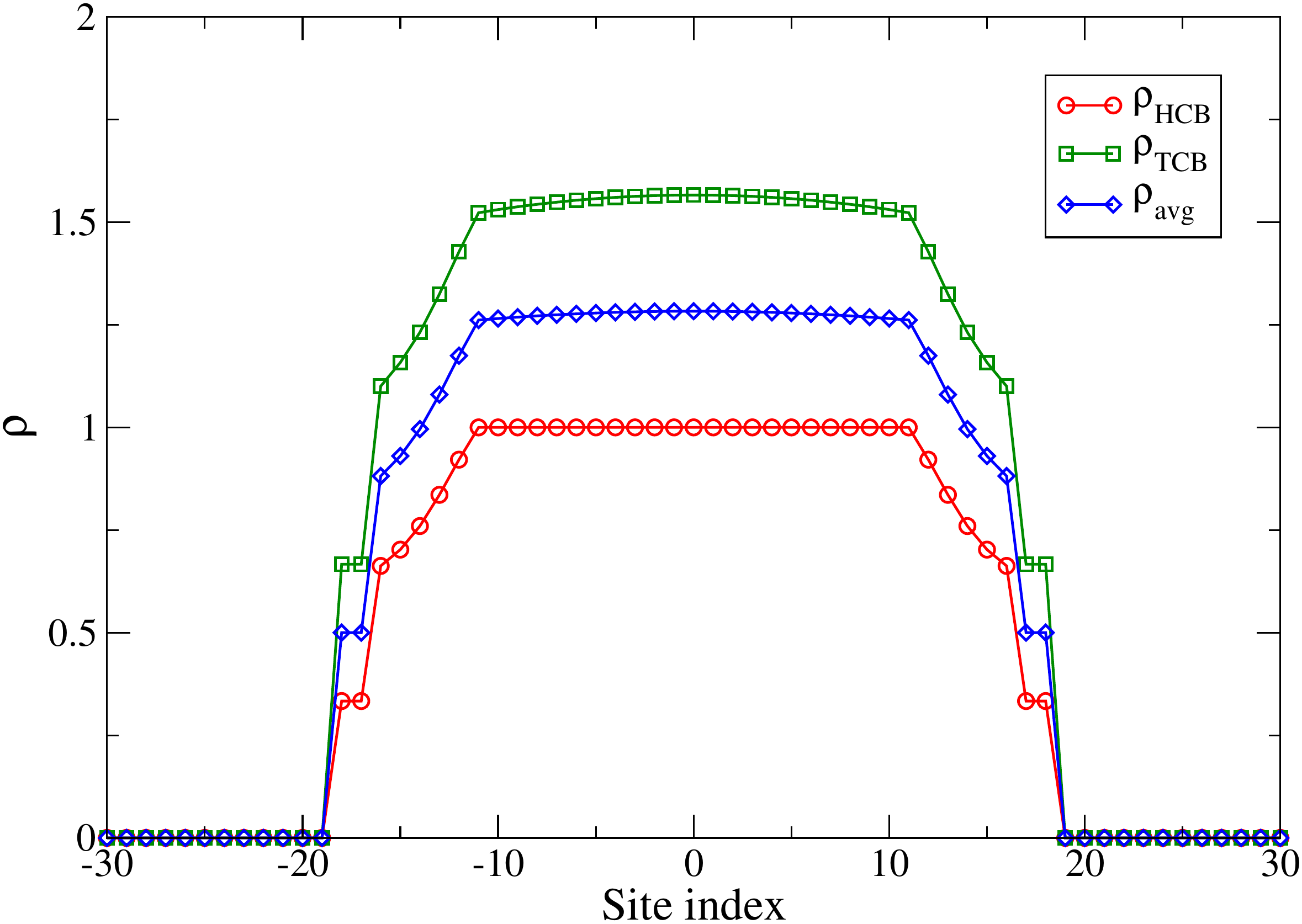}
\caption{(Color online) HCB and TCB confined in harmonic trap with $V_{tr}=0.002$ at $U=0.0$, $t/|V|=0.225$ and $\mu/|V|=-0.6$,
corresponding to Fig.\ref{fig:fig4_HC+3b_t1_U0_Vvar_phase-diagram}.}
\label{fig:trap_fig1}
\end{figure}

\begin{figure}[!t]
   \centering
\includegraphics[width=0.45\textwidth]{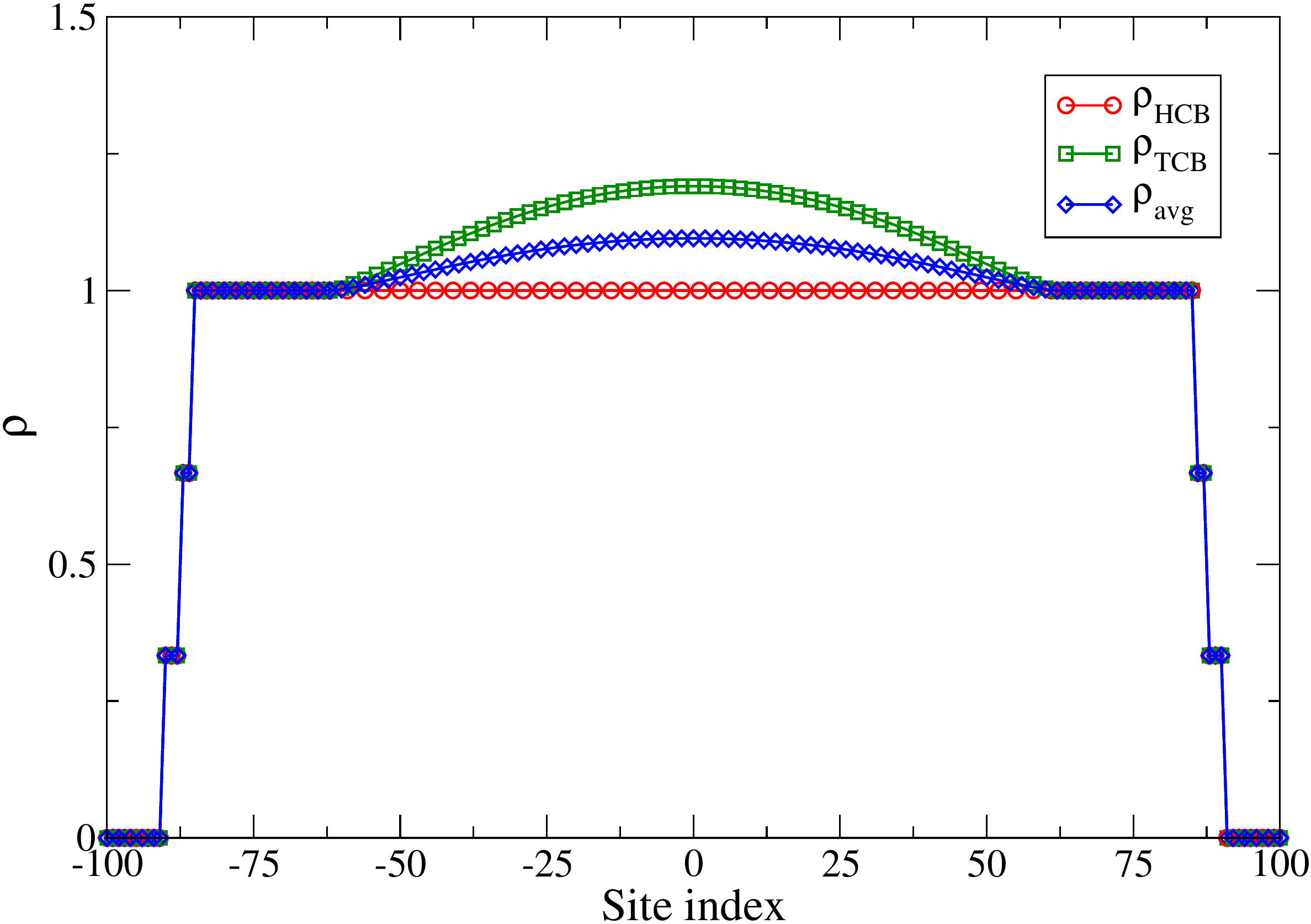}
\caption{(Color online) HCBs and TCBs confined in harmonic trap with $V_{tr}=0.002$ at $|V|/U=0.75$, $t/U=0.1$ and 
$\mu/U=0.0$,
corresponding to the cut shown in Fig.\ref{fig:fig6_HC+3b_U40_V-30_tvar_PD}.}
\label{fig:trap_fig2-2}
\end{figure}

Here we discuss the effect of an external harmonic confinement on the quantum phase diagrams that are investigated in this work.
For this purpose we consider a harmonic trap along the length of the linear chains with equal potential on both the legs and 
repeat the calculations.
An extra term of the form $V_{tr}\sum n x^2$ is thus added to the system Hamiltonian, where $V_{tr}$
is the trap parameter and $x$ is the site index ($=0$ at the center).
This is equivalent to redefining the effective chemical potential as $\mu_x=\mu_0-V_{tr} x^2$. 
For our calculation we set $(V_{tr}/t)=0.002$,  
which is experimentally achievable. 
The results obtained are presented in Figs.~\ref{fig:trap_fig1} and \ref{fig:trap_fig2-2}.
In Fig.~\ref{fig:trap_fig1} we analyze the region along $t/|V|=0.225$ for $U=0.0$ and $\mu_0 /|V|=-0.6$ 
of Fig.~\ref{fig:fig4_HC+3b_t1_U0_Vvar_phase-diagram}.
The result is very similar to the one obtained for the homogeneous case. It is know obvious that the trap center 
has comparatively higher density than other sites away from the center. 
The center of the trap is in the HCBMI+SF phase, which can be clearly seen from the density plot in Fig.~\ref{fig:trap_fig1}.
Moving away from the trap center, 
we obtain a small region of the 2SF phase and then the TSF phase. This can be identified as the jump in the 
density in steps of three particles. 
The overall 
extent of all the phases is close to $40$ sites. The extent of all the phases will increase by considering a shallow trap. 
When we add the same trap to our system with $|V|/U=0.75$, a comparatively larger region of condensate
is seen. This corresponds to a cut along $t/U=0.1$ with 
$\mu_0/U=0.0$. Once again center of the trap exhibits a HCBMI+SF phase.
But unlike the previous case, we first obtain an MI phase as we move away from the center of the trap. Farther going away, it 
yields the PSF phase where the density changes in steps of two particles. At the trap boundaries the particle number vanishes. 
All these phases 
can be probed in the already existing technique of site resolved imaging which has been used to obtain the signature 
of Mott shells in the optical lattice experiments~\cite{kuhr,greiner,kozuma}. 

\section{Conclusions}
\label{sec:sec4}
We have studied the ground state phase diagram of two- and three-body constrained 
dipolar bosons in a system of two linear optical lattices coupled by the dipole-dipole interaction using the self-consistent 
CMFT method. We analyze a wide range of parameters and obtain phase diagrams depicting all the important phases that may arise due to 
the competition between the on-site two-body repulsion and the nearest neighbor attraction with the 
two- and three-body constraints. 
We find that the system exhibits mainly four types of phases; 
namely the 2SF, PSF, TSF and the MI phases corresponding to different ranges of parameters. 
We show the signature of different phases using 
the CMFT calculation which cannot be obtained using the conventional single site mean-field theory. 
By computing the pair correlation
along with the trimer correlation functions, we identify the PSF and TSF phases. 
These results are further substantiated by the the R\'{e}nyi entanglement entropy,
which is found to be finite in the PSF and TSF phases and zero in the MI phase. In addition we discuss the
 effect of external harmonic confining potential, which emphasizes the feasibility of observing these phases 
in an experiment. 

\section{Acknowledgment}
We would like to thank Luis Santos and Sebastian Greschner for many useful discussions.
We acknowledge DST-SERB, India for the financial support through Project No. PDF/2016/000569 and T.M. thanks IIT-Guwahati 
for all the financial supports through the start-up research grant. 
B. K. S. acknowledges financial support from CAS through the PIFI fellowship under the project number 2017VMB0023.
A part of this work was carried out using Vikram-100 computing facilities at 
PRL, Ahmedabad and the remaining works were performed using Param-Ishan HPC facility at IIT-Guwahati.

\end{document}